\documentclass[12pt]{article}
\begin{document}

\title{\bf Plane Symmetric Gravitational Collapse}

\author{M. Sharif \thanks{msharif@math.pu.edu.pk} and Zahid Ahmad
\thanks{zahid$\_$rp@yahoo.com}\\
Department of Mathematics, University of the Punjab,\\
Quaid-e-Azam Campus, Lahore-54590, Pakistan.}

\date{}
\maketitle

\begin{abstract}
In this paper, we derive the general formulation by considering
two arbitrary plane symmetric spacetimes using Israel's method. As
an example, we apply this formulation to known plane symmetric
spacetimes. We take the Taub's static metric in the interior
region whereas Kasner's non-static metric in the exterior region.
It is shown that the plane collapses in some cases whereas it
expands in some other cases.
\end{abstract}


\section{Introduction}

An important issue in relativistic astrophysics and theory of
General Relativity (GR) is to determine the end state of
gravitational collapse. It has many interesting applications in
astrophysics where the formation of compact stellar objects such as
white dwarf and neutron star are formed by a period of gravitational
collapse. The singularity theorems of Hawking and Penrose [1] show
that if a trapped surface forms during the collapse of a compact
object made out of physically reasonable matter, such a collapse
will develop a spacetime singularity. A spacetime singularity is
defined as the region in the spacetime where the evolution of
geodesics will be incomplete.

The singularity theorems do not provide information about the
visibility of a spacetime singularity. A spacetime singularity is
called a naked singularity if it is visible to an observer. A
spacetime singularity which is not visible is called a black hole.
Penrose [2] proposed that the spacetime singularities must be hidden
inside an event horizon. This means that the gravitational collapse
must end as a black hole. This is called the cosmic censorship
conjecture (CCC). The CCC is also one of the most important open
problems in gravitational physics today. Since many physical
applications of black hole and several other areas in GR depend on
CCC, hence a detailed study of dynamically developing gravitational
collapse models is necessary to obtain a correct form of the CCC. To
study gravitational collapse in the framework of GR, it is necessary
to consider the appropriate geometry of interior and exterior
regions and determine proper junction conditions which allow the
matching of these regions.

The pioneering work on gravitational collapse was first started by
Oppenheimer and Snyder [3]. They studied collapse of dust by
considering a static Schwarzschild in exterior and Friedmann like
solution in interior. Since then, many people [4-14] have extended
the above study of collapse by taking an appropriate geometry of
interior and exterior regions. Recently, the effect of a positive
cosmological constant on spherically symmetric collapse with
perfect fluid has been investigated [15] by considering the
matching conditions between static exterior and non-static
interior spacetimes. All these studies are restricted to spherical
gravitational collapse. On the other hand, the study about
non-spherical collapse is limited than spherical collapse. Shapiro
and Teukolsky [16] studied numerically the problem of a dust
spheroid and found that a black hole could be formed if the
spheroid is compact enough. Otherwise, the end state of the
collapse will be a naked singularity. Barrabs et al. [17]
investigated an analytical model of a collapsing convex thin shell
and showed that apparent horizons are not formed in some cases.
These results were soon generalized to more general cases [18].
Since then, these studies have attracted attention to
non-spherical collapse.

Villas da Rocha et al. [19] worked on the self-similar
gravitational collapse of perfect fluid using Israel's method
[20]. Applying the same analysis, Pereira and Wang [21] studied
the gravitational collapse of cylindrical shells made of counter
rotating dust particles. They derived general formulas by
considering two arbitrary cylindrically symmetric spacetimes and
applied these by taking two known cylindrically symmetric
spacetimes. They concluded that in some cases the shell collapses
and in some cases it expands. The same authors [22] have also
discussed the dynamics of expanding and collapsing cylindrically
symmetric fields with lightlike wave-fronts. There is a large body
of literature [23- 26] available on gravitational collapse which
show keen interest to investigate this issue by using Israel's
formulism. In this paper, we extend the work done by Pereira and
Wang [21] for cylindrically symmetric spacetimes to plane
symmetric spacetimes. The paper is outlined as follows. In next
section, we derive general formulae by considering two arbitrary
plane symmetric spacetimes. Section 3 provides application of
these formulae to known plane symmetric spacetimes. We consider
the Taub's static metric in the interior region whereas Kasner's
non-static metric in the exterior region. Finally, in section 4,
we discuss and conclude the results.

\section{General Formalism}

We consider a timelike $3D$ hypersurface $\Sigma$, which divides
$4D$ spacetime into two regions interior and exterior spacetimes,
denoted by $V^+$ and $V^-$ respectively. The region $V^-$ is
described by the metric
\begin{equation}
ds_{-}^2=f^-(t,z)dt^2-g^-(t,z)(dx^2+dy^2)-h^-(t,z)dz^2,
\end{equation}
where $\{\chi^{-\mu}\}\equiv\{t,x,y,z\},~(\mu=0,1,2,3)$ are the
usual Cartesian coordinates. The metric for the region $V^+$ is
given by
\begin{equation}
ds_{+}^2=f^+(T,Z)dT^2-g^+(T,Z)(dx^2+dy^2)-h^+(T,Z)dZ^2,
\end{equation}
where $\{\chi^{+\mu}\}\equiv\{T,x,y,Z\},~(\mu=0,1,2,3)$ is another
set of the Cartesian coordinates.

According to the junction condition [20,27], it is assumed that
the interior and the exterior spacetimes are the same on the
hypersurface $\Sigma$ which can be expressed as
\begin{equation}
(ds_{-}^2)_{\Sigma}=(ds_{+}^2)_{\Sigma}=ds_{\Sigma}^2.
\end{equation}
The extrinsic curvature tensor $K^\pm_{ab}$ to hypersurface
$\Sigma$ is defined as [20]
\begin{equation}
K_{ab}^\pm=n_{\sigma}^\pm(\frac{\partial^2x_\pm^\sigma}{\partial\xi^a\partial\xi^b}
+\Gamma^\sigma_{\mu\nu}\frac{\partial x_\pm^\mu}{\partial\xi^a}
\frac{\partial x_\pm^\nu}{\partial\xi^b}),\quad (\sigma, \mu,
\nu=0,1,2,3),
\end{equation}
where $(a,b=0,1,2)$. Here the Christoffel symbols
$\Gamma^\sigma_{\mu\nu}$ are calculated from the interior or
exterior metrics (1) or (2), $n_{\mu}^\pm$ are the components of
outward unit normals to $\Sigma$ in the coordinates
$\chi^\pm\sigma$. The equations of hypersurface $\Sigma$ in the
coordinates $\chi^\pm\sigma$ are written as
\begin{eqnarray}
k_{-}(z,t)&=&z-z_{0}(t)=0,\\
k_{+}(Z,T)&=&Z-Z_{0}(T)=0.
\end{eqnarray}
Using Eq.(5) in (1), the metric on $\Sigma$ takes the form
\begin{equation}
(ds_{-}^2)_{\Sigma}=[f^-(t,z_0(t))-h^-(t,z_0(t)){z'_0}^2(t)]dt^2-g^-(t,z_0(t))(dx^2+dy^2).
\end{equation}
Similarly, Eqs.(2) and (6) yield
\begin{equation}
(ds_{+}^2)_{\Sigma}=[f^+(T,Z_0(T))-h^+(T,Z_0(T)){Z'_0}^2(T)]dT^2-g^+(T,Z_0(T))(dx^2+dy^2),
\end{equation}
where prime means ordinary differentiation with respect to the
indicated argument. The metric on the hypersurface $\Sigma$ is given
by
\begin{equation}
(ds^2)_{\Sigma}=\gamma_{ab}d\xi^ad\xi^b=d\tau^2-g(\tau)(dx^2+dy^2).
\end{equation}
From the junction condition (3), it follows that
\begin{eqnarray}
d\tau&=&[f^-(t,z_0(t))-h^-(t,z_0(t)){z'_0}^2(t)]^\frac{1}{2}dt,\nonumber\\
&=&[f^+(T,Z_0(T))-h^+(T,Z_0(T)){Z'_0}^2(T)]^\frac{1}{2}dT,\\
g(\tau)&=&g^-(t,z_0(t))=g^+(T,Z_0(T)).
\end{eqnarray}
The outward unit normals in $V^{-}$ and $V^{+}$ can be evaluated by
using Eqs.(5) and (6)
\begin{eqnarray}
n_\mu^+&=&[\frac{f^+h^+}{f^+-h^+{Z'_0}^2(T)}]^\frac{1}{2}(-Z_0'(T),0,0,1),\\
n_\mu^-&=&[\frac{f^-h^-}{f^--h^-{z'_0}^2(t)}]^\frac{1}{2}(-z_0'(t),0,0,1).
\end{eqnarray}
The components of the extrinsic curvature $K_{ab}^\pm$ are
\begin{eqnarray}
K_{\tau\tau}^+&=&-\frac{(f^+h^+)^\frac{1}{2}}
{2[{f^+-h^+{Z'_0}^2(T)}]^\frac{3}{2}}\{-\frac{f^+_{,Z}}
{h^+}+(\frac{f^+_{,T}}{f^+}-2\frac{h^+_{,T}}{h^+})Z_0'(T)\nonumber\\
&+&(2\frac{f^+_{,Z}}{f^+}-\frac{h^+_{,Z}}{h^+}){Z'_0}^2(T)
+\frac{h^+_{,T}}{f^+}{Z'_0}^3(T)-2Z_0''(T)\},\nonumber\\
K_{xx}^+&=&-\frac{1}{2}[\frac{f^+h^+}{f^+-h^+{Z'_0}^2(T)}]^\frac{1}{2}
\{\frac{g^+_{,Z}}{h^+}+\frac{g^+_{,T}}{f^+}Z_0'(T)\}=K_{yy}^+,
\end{eqnarray}
where comma denotes partial derivative and $K_{ab}^-$ can be
obtained from the above expressions by the replacement
\begin{eqnarray}
f^+,g^+,h^+,Z_0(T),T,Z\rightarrow f^-,g^-,h^-,z_0(t),t,z.
\end{eqnarray}
The surface energy-momentum tensor, $\tau_{ab}$, is defined as
[20,21]
\begin{equation}
\tau_{ab}=\frac{1}{\kappa}\{[K_{ab}]^--\gamma_{ab}[K]^-\},
\end{equation}
where $\kappa$ is the gravitational constant and
\begin{equation}
[K_{ab}]^-=K_{ab}^+-K_{ab}^-,\quad [K]^-=\gamma_{ab}[K_{ab}]^-.
\end{equation}
Using Eq.(14) and the corresponding expressions for $K_{ab}^-$ into
Eq.(16), $\tau_{ab}$ can be expressed in the form
\begin{equation}
\tau_{ab}=\rho w_aw_b+p(x_ax_b+y_ay_b),\quad(a,b=\tau,x,y),
\end{equation}
where $\rho$ is the surface energy density, $p$ is the tangential
pressure provided that they satisfy some energy conditions [28]
and $w_a,~x_a,~y_a$ are unit vectors defined on the surface given
by
\begin{equation}
w_a=\delta^\tau_a, \quad x_a=h^\frac{1}{2}(\tau)\delta^x_a,\quad
y_a=h^\frac{1}{2}(\tau)\delta^y_a.
\end{equation}
Here $\rho$ and $p$ turn out to be
\begin{eqnarray}
\rho=\frac{2}{\kappa g(\tau)}[K_{xx}]^-,\quad
p=\frac{1}{\kappa}[[K_{\tau\tau}]^--\frac{[K_{xx}]^-}{g(\tau)}].
\end{eqnarray}

\section{Plane Symmetric Gravitational Collapse}

In this section, we shall apply the general results developed in
the previous section by considering known plane symmetric
spacetimes. The metric for interior region (Taub's static metric)
is given by [29]
\begin{equation}
ds_{-}^2=z^\frac{-1}{2}(dt^2-dz^2)-z(dx^2+dy^2),\quad z>0.
\end{equation}
For exterior spacetime, we take the non-static (Kasner's) metric
given by [29]
\begin{equation}
ds_{+}^2=T^\frac{-1}{2}(dT^2-dZ^2)-T(dx^2+dy^2),\quad T>0.
\end{equation}
Both interior and exterior spacetimes are vacuum, i.e.,
energy-momentum for both spacetimes is zero. The physical
importance to consider these (Taub and Kasner) metrics is that
they satisfy energy conditions [28]. Using Eqs.(21) and (22), the
junction conditions (10) and (11) take the form
\begin{eqnarray}
d\tau&=&[1-{z'_0}^2(t)]^\frac{1}{2}z_0^\frac{-1}{4}dt
=[1-{Z'_0}^2(T)]^\frac{1}{2}T^\frac{-1}{4}dT,\\
z_{0}(t)&=&T.
\end{eqnarray}
From Eqs.(23) and (24), it turn out that
\begin{equation}
(\frac{dT}{dt})^2=\frac{1}{\Delta^2}\equiv[2-{Z'_0}^2]^{-1}
\end{equation}
and Eqs.(24) and (25) yield
\begin{equation}
z''_0(t)=\frac{d^2T}{dt^2}=\frac{Z'_0Z''_0}{\Delta^4}.
\end{equation}
Using Eq.(14) and the corresponding expressions for $K_{ab}^-$
into Eq.(20) and considering Eqs.(24)-(26), we obtain
\begin{eqnarray}
\rho&=&\frac{2}{\kappa T^\frac{3}{4}(1-{Z'_0}^2)^\frac{1}{2}}(\Delta-Z'_0),\\
p&=&\frac{\Delta-Z'_0}{4\kappa
T^\frac{3}{4}\Delta(1-{Z'_0}^2)^\frac{3}{2}} \{4T
Z''_0-\Delta({1-Z'_0}^2)\}.
\end{eqnarray}
In order to see the minimal effects of the plane on the collapse,
we shall take $p=0$ in Eq.(28). This leads to the following two
cases
\begin{eqnarray}
(A)&:&\quad\Delta-Z'_0=0, \\
or\quad (B)&:&\quad 4T Z''_0-\Delta({1-Z'_0}^2)=0,
\end{eqnarray}
where
\begin{equation}
\Delta=\pm\sqrt{2-{Z'_0}^2}.
\end{equation}
It is mentioned here that equation of state i.e., $\rho=ap$ can be
used to see pressure effects on the gravitational collapse. Since,
we want to see the minimum effects of the shell on the collapse
therefore, we take $p=0$. The case(A) is trivial because surface
energy density becomes zero from Eq.(27) and thus we leave it.
Case(B) is further divided into two subcases:\\\\
B(i)\quad$4T Z''_0-\sqrt{2-{Z'_0}^2}({1-Z'_0}^2)=0$,\\
B(ii)\quad $4T Z''_0+\sqrt{2-{Z'_0}^2}({1-Z'_0}^2)=0.$\\\\
In the following, we solve these two cases separately.

For the case B(i), integration of the equation
\begin{equation}
4T Z''_0-\sqrt{2-{Z'_0}^2}({1-Z'_0}^2)=0
\end{equation}
yields
\begin{equation}
Z'_0(T)=\pm\frac{\sqrt{2}\tanh(\frac{\ln(T)}{4})}{\sqrt{1+\tanh(\frac{\ln(T)}{4})^2}},
\end{equation}
where constant of integration is taken zero for the sake of
simplicity.  It is mentioned here that the $"+"$ and $"-"$ signs
correspond to an expanding and
collapsing planes respectively.\\
\par\noindent
\textbf{Expanding plane}\\
\par\noindent
Integrating Eq.(33) with "+" sign, we obtain
\begin{equation}
Z_0(T)=\frac{(-2+\sqrt{T}-2T+T^\frac{3}{2}-\sqrt{1+T}\sinh^{-1}[\sqrt{T}])}
{(T^\frac{1}{4}\sqrt{\frac{1+T}{\sqrt{T}}})},
\end{equation}
where integration constant is taken zero to avoid complicated
situation. Eqs.(33) and (34) yield respectively
\begin{eqnarray}
Z_0(T)&=&\left \{ \begin{array}{lll}
0.126418&,&T=7,\nonumber\\
+\infty&,&T=+\infty,
\end{array}\right.\nonumber\\
Z'_0(T)&=&\left \{ \begin{array}{lll}
0.581861&,&T=7,\\
1&,&T=+\infty
\end{array}\right.
\end{eqnarray}
indicating that the plane is expanding. The plane starts to expand
at time $T=7$, where it has displacement $0.126418$, velocity
$0.581861$ and positive acceleration. The expansion ends at
$T=+\infty$, where the plane has displacement $+\infty$, velocity
unity and zero acceleration. It can be seen from Eq.(27) that the
energy density decreases from finite value to zero in this
interval.
\newpage
\par\noindent
\textbf{Collapsing plane}\\
\par\noindent
Integration of Eq.(33) with "-" sign yields
\begin{equation}
Z_0(T)=-\frac{(-2+\sqrt{T}-2T+T^\frac{3}{2}-\sqrt{1+T}\sinh^{-1}[\sqrt{T}])}
{(T^\frac{1}{4}\sqrt{\frac{1+T}{\sqrt{T}}})},
\end{equation}
where integration constant is taken zero. From Eqs.(33) and (36),
we obtain collapsing plane
\begin{eqnarray}
Z_0(T)&=&\left \{ \begin{array}{lll}
2.29559&,&T=1,\nonumber\\
0.439096&,&T=6,
\end{array}\right.\nonumber\\
Z'_0(T)&=&\left \{ \begin{array}{lll}
0&,&T=1,\\
-0.547856&,&T=6.
\end{array}\right.
\end{eqnarray}
The plane starts to collapse at time $T=1$, where it has radial
displacement $2.29559$, zero velocity and positive acceleration and
ends at $T=6$, where the plane has displacement 0.439096 and
negative velocity.  It is to be noticed that energy density remains
finite in this interval i.e., collapse does end as a singularity.
This situation might be change if one replaces interior and exterior
metrics which are physical i.e., which satisfy energy conditions. It
is mentioned here that plane collapses and expands along
$Z$-direction only because $Z$ is not a radial coordinate.

The case B(ii) gives similar results as the case B(i).

\section{Conclusion}

This paper is an extension of the previous studies of spherical
and cylindrical gravitational collapse to plane symmetric
gravitational collapse using Israel's method. First of all, we
have developed a general formalism for two arbitrary plane
symmetric spacetimes in terms of the metric coefficients and their
first derivatives. Then, as an example, we have applied this
formulation to known plane symmetric spacetimes. We have taken the
Taub's static metric in the interior region whereas Kasner's
non-static metric in the exterior region. We have taken these
metrics because they satisfy energy conditions [28]. It is found
that the plane collapses in some cases whereas it expands in some
other cases.

There arise two main cases $A$ and $B$ according to Eqs.(29) and
(30) respectively. The case $A$ is trivial. For the case $B$,
there is expanding as well as collapsing process. It is mentioned
here that in one case of B(i), the plane collapses for all $T>0$
and in other case of B(i), it expands for all $T>0$. But
physically, the plane collapses in the interval $1\leq T\leq 6$
and expands in the interval $7\leq T<\infty $. Notice that the
collapsing interval is less than the expanding interval. Since we
have taken vacuum spacetimes in the interior and exterior regions.
This result is consistent with that the cosmological constant,
i.e., vacuum energy density slows down the collapsing process
[30].

We would like to mention here that the collapse does not end as a
singularity because energy density remains finite at all times in
both cases. The physical implication of this work is that the
general formalism can be applied by considering different physical
plane symmetric spacetimes (which satisfy energy conditions) to get
interesting result.

\vspace{0.5cm}

{\bf Acknowledgment}

\vspace{0.5cm}

We appreciate the Higher Education Commission Islamabad, Pakistan,
for its financial support during this work through the {\it
Indigenous PhD 5000 Fellowship Program Batch-I}.

\vspace{0.5cm}

{\bf References}

\begin{description}

\item{[1]} Penrose, R.: Phys. Rev. Lett. {\bf 14}(1965)57;  Hawking, S.W.:
{\it Proc.R. Soc. London} {\bf A300}(1967)187; Hawking, S.W. and Penrose,
R.: {\it Proc. R. Soc. London} {\bf A314}(1970)529.

\item{[2]} Penrose, R.: Riv. Nuovo Cimento {\bf 1}(1969)252.

\item{[3]} Oppenheimer, J.R. and Snyder, H.: Phys. Rev. {\bf 56}(1939)455.

\item{[4]} Misner, C.W. and Sharp, D.: Phys. Rev. {\bf 136}(1964)b571.

\item{[5]} Ghosh, S.G. and Deshkar, D.W.: Int. J. Mod. Phys. {\bf D12}(2003)317.

\item{[6]} Ghosh, S.G. and Deshkar, D.W.: Gravitation and Cosmology {\bf 6}(2000)1.

\item{[7]} Debnath, U., Nath, S. and Chakraborty, S.: Mon. Not. R. Astron.
Soc. {\bf 369}(2006)1961.

\item{[8]} Debnath, U., Nath, S. and Chakraborty, S.: Gen. Relativ. Grav. {\bf 37}(2005)215.

\item{[9]} Smith, W.L. and Mann, R.B.: Phys. Rev. {\bf D56}(1997)4942.

\item{[10]} Ross, S.F. and Mann, R.B.: Phys. Rev. {\bf D47}(1993)3319.

\item{[11]} Mann, R.B. and Oh, J.J.: Phys. Rev. {\bf D74}(2006)124016.

\item{[12]} Madhav, T.A., Goswami, R. and Joshi, P.S.: Phys. Rev. {\bf
D72}(2005)084029

\item{[13]} Ilha, A., Kleber, A. and Lemos, J.P.S.: J. Math.Phys. {\bf
40}(1999)3509.

\item{[14]} Maharaj, S.D. and Govender, M.: Pramana J. Phys. {\bf
54}(2000)715.

\item{[15]} Sharif, M. and Ahmad, Z.: Mod. Phys. Lett. \textbf{A22}
(2007)1493.

\item{[16]} Shapiro, S.L. and Teukolsky, S.A.: Phys. Rev. Lett. {\bf 66}(1991)994.

\item{[17]} Barrabes, C., Israel, W. and Letelier, P.S.: Phys. Lett. {\bf A160}(1991)41.

\item{[18]} Barrabes, C., Gramain, A., Lesigne, E. and Letelier, P.S.:
Class. Quantum Grav. {\bf 9}(1992)L105.

\item{[19]} Villas da Rocha, J.F., Wang, A. and Santos, N.O.: Phys. Lett. {\bf A255}(1999)213.

\item{[20]} Israel, W.: Nuovo Cimento {\bf B44}(1966)1; {\it
ibid} {\bf B48}(1967)463(E).

\item{[21]} Pereira, P.R.C.T. and Wang, A.: Phys. Rev. {\bf
D62}(2000)124001; Erratum {\it ibid} {\bf D67}(2003)129902.

\item{[22]} Pereira, P.R.C.T. and Wang, A.: Int. J. Mod. Phys. {\bf D11}(2002)561.

\item{[23]} Lemos, J.P.S.: Phys. Rev. {\bf D57}(1998)4600.

\item{[24]} Cai, R.G. and Wang, A.: Phys. Rev. {\bf D73}(2006)063005.

\item{[25]} Herrera, L. and Santos, N.O.: Phys. Rev. {\bf D70}(2004)084004.

\item{[26]} Herrera, L. and Santos, N.O.: Class. Quant. Grav. {\bf 22}(2005)2407.

\item{[27]} Santos, N.O.: Mon. Not. R. Astron.
Soc. {\bf 216}(1985)403.

\item{[28]} Hawking, S.W. and Ellis, G.F.R: {\it The Large Scale Structure of Spacetime}
(Cambridge University Press, Cambridge, 1973).

\item{[29]} Stephani, H., Kramer, D., Maccallum, M., Hoenselaers, C. and Herlt, E.:
 {\it Exact Solutions of Einstein's Field Equations} (Cambridge University Press, Cambridge, 2003).

\item{[30]} Cissoko, M., Fabris, J.C., Gariel, J., Denmat, G.L. and Santos, N.O.:
 arXiv:gr-qc/9809057.
\end{description}
\end{document}